\documentclass[a4paper]{jpconf}

\usepackage{graphicx}
\RequirePackage[dvipsnames]{xcolor}
\usepackage[square,sort,comma,numbers]{natbib}        % required for bibliography

\usepackage{subcaption}

\usepackage{amsmath} % assumes amsmath package installed
\usepackage{amssymb}  % assumes amsmath package installed

\renewcommand\theenumi{\arabic{enumi}}

\newcommand{\R}{\mathbb{R}}
\newcommand{\K}{\mathcal{K}}
\newcommand{\bk}{\mathbf{k}}
\newcommand{\E}{\mathcal{E}}
\newcommand{\cut}{\text{cut}}
\newcommand\red[1]{{\color{red}#1}}
\newcommand\green[1]{{\color{OliveGreen}#1}}

\begin{document}
\title{Translating Assembly Accuracy Requirements to Cut-Off Frequencies for Component Mode Synthesis}

\author{Lars A.L. Janssen$^1$, Bart Besselink$^2$, Rob H.B. Fey$^1$, and Nathan van de Wouw$^1$}
\address{$^1$ Dynamics \& Control, Department of Mechanical Engineering,  Eindhoven University of Technology, $^2$  Bernoulli Institute for Mathematics, Computer Science and Artificial Intelligence, University of Groningen}
\ead{l.a.l.janssen@tue.nl}

\begin{abstract}
One of the most popular methods for reducing the complexity of assemblies of finite element models in the field of structural dynamics is component mode synthesis.
A main challenge of component mode synthesis is balancing model complexity and model accuracy, because it is difficult to predict how component reduction influences assembly model accuracy.
This work introduces an approach that allows for the translation of assembly model accuracy requirements in the frequency domain to the automatic selection of the cut-off frequencies for the model-order reduction (MOR) of components.
The approach is based on a mathematical approach for MOR for coupled linear systems in the field of systems and control.
We show how this approach is also applicable to structural dynamics models.
We demonstrate the use of this approach in the scope of component mode synthesis (CMS) methods with the aim to reduce the complexity of component models while guaranteeing accuracy requirements of the assembly model. 
The proposed approach is illustrated on a mechanical, three-component structural dynamics system for which reduced-order models are computed that are reduced further compared to reduction using standard methods. 
This results in lower simulation cost, while maintaining the required accuracy.
\end{abstract}

\section{INTRODUCTION}
To predict, analyze, optimize and control the behavior of dynamic systems, models are required that 1) accurately represent the system dynamics and 2) can be used with limited computational cost.
Generally, this results in a trade-off between the accuracy and complexity of the model.
The goal of model order reduction methods is to manage this trade-off.
The field of MOR spans several disciplinary fields, such as structural dynamics, numerical mathematics, and systems and control \cite{besselink2013}.
Examples from these fields are modal truncation methods~\cite{hintz1975analytical}, balancing methods~\citep{moore1981}, and Krylov methods~\citep{grimme1997}, respectively, which have in common that they rely on projections.

In these fields, complex systems often consist of multiple interacting components, which makes managing the complexity especially challenging.
This is a result of the fact that models of interconnected subsystems are not only required to be accurate and computationally efficient, but they also need to preserve the interconnection structure.
Furthermore, since individual components are often developed by independent teams, the process of finding a reduced-order model (ROM) is typically also done on component-level.
Note that in the field of systems and control, such methods are referred to as modular or subsystem MOR \cite{vaz1990,lutowska2012}.
However, in the field of structural dynamics, the most popular component-level approach is component mode synthesis, which has been studied since the early sixties \cite{hurty1960vibrations} and remains an active area of research to this date \cite{allen2019substructuring,kessels2022sensitivity}.
CMS is also a projection-based approach that typically involves identifying and extracting the most important vibration modes from the component Finite Element (FE) models and then combining them in a way that accurately represents the dynamic behavior of the original assembly. 
There are several different CMS methods, such as the Craig-Bampton, Hintz-Herting, and Rubin CMS-methods \cite{craig2000coupling,geradin2014mechanical}. 

For all of these methods, the complexity of component models is reduced individually, which generally leads to a loss of accuracy of the ROM of the component in comparison to the original higher-order component model.
When these reduced component models are assembled, the errors introduced by the component reduction could potentially accumulate and lead to a poor accuracy of the reduced assembly model. 
With CMS, there are typically no a priori error bounds available that can assist in the prediction of the propagation of component-level reduction errors to the assembly model.
Therefore, there is a need for methods that can translate requirements on the accuracy of the assembly model to the individual component models.

Typically, accuracy requirements on assembly level take the form of a frequency range of interest, in which the dynamics should be accurately represented.
Each of the components is then reduced by selecting the component modes up to a predefined cut-off frequency which is typically chosen to be two or three times the maximum frequency of interest for the assembly \cite{voormeeren2012dynamic, seshu1997substructuring}.
However, this standard approach does not take explicitly into account how component reduction errors propagate to assembly level. 
As a result, a reduced-order component could be reduced too conservatively, i.e., not enough, such that the computational cost of using the ROM is unnecessarily high.
However, it may also be possible that a reduced-order component could be reduced too much, such that the assembly level specification is not met. 

In earlier work \cite{janssen2022modular,janssen2023modular}, we have developed a framework for relating a priori component-level reduction errors to reduction errors on assembly level in the frequency domain. 
This is achieved by casting this relation as a robust performance analysis problem from the field of robust control \cite{zhou1998}. 
Specifically, this framework can be used to compute component-level accuracy specifications directly based on accuracy specifications for the assembly.

The main contribution of the current paper is to employ this framework for component mode synthesis for structural dynamics models. 
In particular, this allows for the computation of component-level cut-off frequencies for the selection of component modes, in such a way that accuracy requirements on the assembly model are guaranteed to be satisfied.
%This is achieved with a reformulation of the established approach \cite{janssen2022modular,janssen2023modular} such that second-order ordinary differential equations (ODE) can be directly applied. 
We apply this approach using the Hintz-Herting CMS-reduction method to find reduced-order models and show that a further reduction of component models in comparison to typical methods can be achieved while guaranteeing the required accuracy on the assembly model.
It is important to mention, however, that our approach can be applied to any other CMS method as well.

Furthermore, this approach is illustrated on a three-component structural dynamics system and compared with the standard approach of a priori selecting a fixed cut-off frequency for all components.
We show that with our new method, one of the components can be reduced significantly more compared to the standard reduction approach, resulting in a significantly lower cost of simulation, while the other two components instead need to be slightly more accurate to meet the requirements of the assembly ROM.
This methodology allows engineering teams responsible for component models and designs to make well-informed, collaborative decisions on component model reduction from the perspective of the total assembly.

The paper is organized as follows. Section~\ref{sec:framework} gives the modelling framework from component discretization and reduction to assembly coupling. 
In Section~\ref{sec:methodology}, we describe the proposed general approach to relate assembly requirements to component requirements for CMS.
This approach is demonstrated on an illustrative structural dynamics example system in Section~\ref{sec:example}.
Finally, the conclusions are given in Section~\ref{sec:conclusions}.
\begin{figure}
  	\centering
   	\includegraphics[scale=1, page=1]{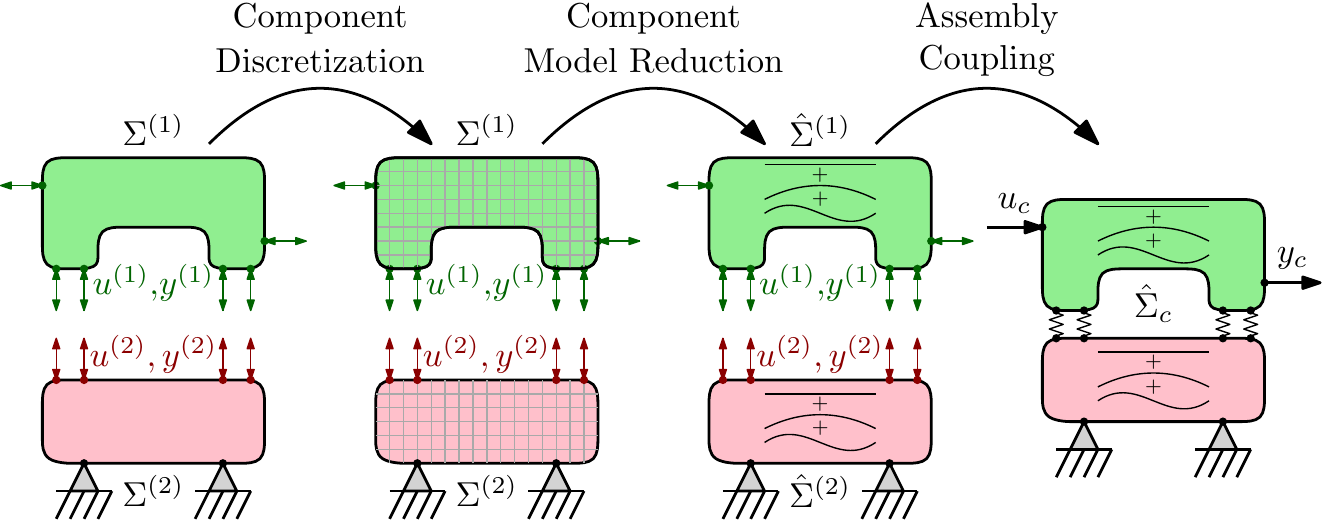}
   	\caption{General approach for model order reduction of a system consisting of multiple components.}
    \label{fig:gen_approach}
\end{figure}
\section{Component-based model reduction}
\label{sec:framework}
A general approach to derive an assembly model that is suitable for applications such as dynamical analysis, simulation, or control, as illustrated in Figure~\ref{fig:gen_approach}, is elaborated in this section.
\subsection{Component Discretization}
Consider a set of $k$ components for which, when spatially discretized, each component $j \in \bk:=\{1,\dots,k\}$ can be represented by a high-order, linear, time-invariant, second-order (FE) model:
\begin{equation}
\label{eq:sigma_j}
\Sigma^{(j)} : \left\lbrace M^{(j)}\ddot{q}^{(j)} + C^{(j)} \dot{q}^{(j)} + K^{(j)}q^{(j)} = B^{(j)} u^{(j)}, \quad y^{(j)} = F^{(j)} q^{(j)}. \right.
\end{equation}
Here, the $n_j$ degrees of freedom (DOF) are given by $q^{(j)}$, and $M^{(j)}, C^{(j)}$, and $K^{(j)}$ represent the component's mass, damping, and stiffness matrix, respectively, $B^{(j)}$ is the input matrix, and $F^{(j)}$ is the output matrix for displacement. 
Furthermore, denote by $H^{(j)}(i\omega)$ the frequency response function (FRF) from the $m_j$ input forces $u^{(j)}$ to the $p_j$ output positions $y^{(j)}$ given as
\begin{equation}
\label{eq:H}
H^{(j)}(i\omega) := F^{(j)} (-\omega^2 M^{(j)} + i\omega C^{(j)} + K^{(j)} )^{-1} B^{(j)}.
\end{equation}
%H^{(j)}(\omega) := \left[\begin{array}{cc} 
%\hspace{-5pt}F^{(j)}\hspace{-5pt} & 0\hspace{-5pt}
%\end{array}\right] \left(i \omega I  - \left[\begin{array}{cc}
%0 & I \\ -(M^{(j)})^{-1}K^{(j)} & -(M^{(j)})^{-1} C^{(j)}
%\end{array}\right] \right)^{-1}\left[\begin{array}{c}
%\hspace{-5pt}0\hspace{-5pt} \\ \hspace{-5pt}(M^{(j)})^{-1}B^{(j)}\hspace{-5pt}
%\end{array}\right].
Note that alternative methods are available to compute this FRF using modal analysis \cite{de2004numerical,geradin2014mechanical}.

\subsection{Component Model Reduction}
For complex components, the number of DOF $n_j$ is generally too large for parameter studies, design optimization, and control applications. 
Therefore, the need arises to compute a reduced-order model $\hat{\Sigma}^{(j)}$ with $\hat{n}_j \ll n_j$ DOF $\hat{q}^{(j)}$.
A popular family of methods to reduce the dimensions of interconnected component models is component mode synthesis (CMS) methods \cite{craig2000coupling}.
In general, the component DOF $q^{(j)}$ are partitioned, i.e., ${q^{(j)}}^\top = \left[\begin{array}{cc} {q^{(j)}_i}^\top & {q^{(j)}_b}^\top\end{array}\right]$, where $q_i$ are internal and $q_b$ are boundary DOF. Boundary DOF are (user specified) DOF which are loaded by external loads and/or adjacent components.
The matrices in (\ref{eq:sigma_j}) can be partitioned accordingly, leading to
\begin{align}
M^{(j)}= \left[\begin{array}{cc}
\hspace{-5pt}M^{(j)}_{ii}\hspace{-5pt} & \hspace{-5pt}M^{(j)}_{ib}\hspace{-5pt} \\ \hspace{-5pt}M^{(j)}_{bi}\hspace{-5pt} & \hspace{-5pt}M^{(j)}_{bb}\hspace{-5pt} \end{array}\right]\hspace{-3pt},
C^{(j)} = \left[\begin{array}{cc}
\hspace{-5pt}C^{(j)}_{ii}\hspace{-5pt} & \hspace{-5pt}C^{(j)}_{ib}\hspace{-5pt} \\ \hspace{-5pt}C^{(j)}_{bi}\hspace{-5pt} & \hspace{-5pt}C^{(j)}_{bb}\hspace{-5pt} \end{array}\right]\hspace{-3pt},
K^{(j)} = \left[\begin{array}{cc}
\hspace{-5pt}K^{(j)}_{ii}\hspace{-5pt} & \hspace{-5pt}K^{(j)}_{ib}\hspace{-5pt} \\ \hspace{-5pt}K^{(j)}_{bi}\hspace{-5pt} & \hspace{-5pt}K^{(j)}_{bb}\hspace{-5pt} \end{array}\right]\hspace{-3pt},
B^{(j)}{=}\left[\begin{array}{c} \hspace{-5pt}O\hspace{-5pt} \\\hspace{-5pt}B^{(j)}_{b}\hspace{-5pt}\end{array} \right]\hspace{-3pt},
{F^{(j)}}^\top{=}\left[\begin{array}{c} \hspace{-5pt}O\hspace{-5pt} \\ \hspace{-5pt}{F^{(j)}_{b}}^\top\hspace{-5pt}\end{array} \right] \hspace{-3pt}.
\end{align} 
%For a more comprehensive overview of these component mode synthesis methods, see \cite{craig2000coupling}.
%In general, each of these methods can be applied to the method that will be proposed.
In this work, the \emph{Hintz-Herting (HH)} \cite{herting1985general} CMS method is applied, although our methodology can be applied to any CMS method. 
In short, in the HH CMS method, component ROMs are computed using the following approach:
\begin{enumerate}
\item The \emph{static constraint modes} $\Psi^{(j)} := -(K_{ii}^{(j)})^{-1}K^{(j)}_{ib}$ are computed. 
\item The eigenvalue problem $(K^{(j)}-\omega^2_\ell M^{(j)})\phi^{(j)}_\ell = 0$ is partly solved for eigenfrequencies up to a user-defined cut-off frequency, i.e., $0 \leq \omega_\ell \leq  \omega_{\cut}$. 
These eigenmodes $\phi^{(j)}_\ell$ are split in rigid body modes ($\omega_\ell = 0$), if any, and elastic free-interface eigenmodes ($0 < \omega_\ell \leq \omega_{\cut}$), and, respectively, collected in matrices ${\Phi_r^{(j)}}^\top = \left[\begin{array}{cc} {\Phi_{r,i}^{(j)}}^\top & {\Phi_{r,b}^{(j)}}^\top \end{array}\right]$ and $
{\Phi_e^{(j)}}^\top = \left[\begin{array}{cc} {\Phi_{e,i}^{(j)}}^\top & {\Phi_{e,b}^{(j)}}^\top \end{array}\right],$
where the matrices $\Phi_r$ and $\Phi_e$ are partitioned according to the internal and boundary DOF.
\item The \emph{uncoupled (from $q_b$) elastic eigenmodes} $\Phi_{ue}^{(j)} := \Phi_{e,i}^{(j)} - \Psi^{(j)}\Phi_{e,b}^{(j)}$ are computed.
\item The \emph{inertia relief modes} $\Phi_{ir}^{(j)} := -(K_{ii}^{(j)})^{-1}(M_{ib}^{(j)}+M_{ii}^{(j)}\Psi^{(j)})\Phi_{r,b}^{(j)}$ are computed. Note that the number of inertia relief modes equals the rigid body modes and may be zero.
\item The reduction matrix for the HH reduction method is then given by
\begin{align}
\label{eq:T}
T^{(j)} &:= \left[\begin{array}{ccc}
\Phi_{ir}^{(j)} & \Phi_{ue}^{(j)} & \Psi^{(j)} \\
O & O & I
\end{array}\right].
\end{align}
\item Using (\ref{eq:T}), the component system matrices (\ref{eq:sigma_j}) can be reduced to obtain
\begin{align*}
\hat{M}^{(j)} &= {T^{(j)}}^\top M^{(j)} T^{(j)},& \hat{K}^{(j)} &= {T^{(j)}}^\top K^{(j)} T^{(j)},& \hat{C}^{(j)} &= {T^{(j)}}^\top C^{(j)} T^{(j)}, \\
\hat{B}^{(j)} &= {T^{(j)}}^\top B^{(j)},& \hat{F}^{(j)} &= F^{(j)} T^{(j)},
\end{align*}
and the reduced-order component model for $j \in \bk$ becomes
\begin{equation}
\label{eq:hatsigma_j}
\hat{\Sigma}^{(j)} : \left\lbrace \hat{M}^{(j)}\ddot{\hat{q}}^{(j)} + \hat{C}^{(j)} \dot{\hat{q}}^{(j)} + \hat{K}^{(j)}\hat{q}^{(j)} = B^{(j)} \hat{u}^{(j)} , \quad \hat{y}^{(j)} = \hat{F}^{(j)} \hat{q}^{(j)}. \right.
\end{equation} 
\end{enumerate}
Generally, the reduction of the component model $\Sigma^{(j)}$ to $\hat{\Sigma}^{(j)}$ introduces a loss of accuracy.
To quantify this accuracy loss, after denoting the FRF of (\ref{eq:hatsigma_j}) by $\hat{H}^{(j)}(i\omega)$ as in (\ref{eq:H}), we introduce the component's error dynamics, for $\omega\in\R$, as
$E^{(j)}(i\omega) := \hat{H}^{(j)}(i\omega) - H^{(j)}(i\omega)$.

\subsection{Assembly Coupling}
In this work, we assume that components are connected through elastic (but potentially very stiff) interfaces.
For all the high-order component models, the input forces are combined into $u_b^\top = [{u^{(1)}}^\top,\dots,{u^{(k)}}^\top]$ and the output positions into $y_b^\top = [{y^{(1)}}^\top,\dots,{y^{(k)}}^\top]$, and $\hat{u}_b$ and $\hat{y}_b$ are combined similarly for the component ROMs.
Both the high-order assembly model $\Sigma_c$ and the reduced-order assembly model $\hat{\Sigma}_c$ are then formed by connecting the components $\Sigma^{(j)}$ as in (\ref{eq:sigma_j}) and $\hat{\Sigma}^{(j)}$ as in (\ref{eq:hatsigma_j}) for all $j \in \bk$, respectively, through the interconnection matrix $\K$, as given by
\begin{align}
\label{eq:sigmac}
\left[\begin{array}{c}
u_b \\ y_c 
\end{array}\right] = 
\underbrace{\left[\begin{array}{cc}
\K_{bb} & \K_{bc} \\
\K_{cb} & O
\end{array}\right]}_{=\K} \left[\begin{array}{c}
y_b \\ u_c 
\end{array}\right] \quad \text{and} \quad \left[\begin{array}{c}
\hat{u}_b \\ \hat{y}_c 
\end{array}\right] = 
\underbrace{\left[\begin{array}{cc}
\K_{bb} & \K_{bc} \\
\K_{cb} & O
\end{array}\right]}_{=\K} \left[\begin{array}{c}
\hat{y}_b \\ u_c 
\end{array}\right],
\end{align}
respectively.
Here, $u_c$ are all $m_c$ external input forces, and $y_c$ and $\hat{y}_c$ are all $p_c$ external output positions of $\Sigma_c$ and $\hat{\Sigma}_c$, respectively, and the total number of DOF are $n_c := \sum_{j=1}^k n_j$ and $\hat{n}_c := \sum_{j=1}^k \hat{n}_j$ in $\Sigma_c$ and $\hat{\Sigma}_c$, respectively.
Given $H_b := \text{diag}(H^{(1)},\dots,H^{(k)})$ and $\hat{H}_b := \text{diag}(\hat{H}^{(1)},\dots,\hat{H}^{(k)})$, the FRFs from $u_c$ to $y_c$ and from $u_c$ to $\hat{y}_c$ are  
\begin{equation}
\label{eq:Gc}
H_c(i\omega) := \K_{cb}H_b(i\omega)( I - \K_{bb}H_b(i\omega))^{-1}\K_{bc} \text{ and } \hat{H}_c(i\omega) := \K_{cb}\hat{H}_b(i\omega)( I - \K_{bb}\hat{H}_b(i\omega))^{-1}\K_{bc},
\end{equation}
respectively, and the assembly model \emph{error dynamics} are then given by
\begin{align}
E_c(i\omega) := \hat{H_c}(i\omega){-}H_c(i\omega) = \K_{cb}\left(\hat{H}_b(i\omega)( I{-}\K_{bb}\hat{H}_b(i\omega))^{-1}{-}H_b(i\omega)( I{-}\K_{bb}H_b(i\omega))^{-1}\right)\K_{bc}
\end{align}
for $\omega \in \R$, i.e., the difference between the high-order and reduced-order assembly model FRFs.

\section{METHODOLOGY FOR ASSEMBLY-AWARE COMPONENT REDUCTION}
\label{sec:methodology}
Typically, in MOR, the aim is to find a reduced-order \emph{assembly} model that is sufficiently accurate in a certain frequency range.
However, with CMS methods, components are reduced individually. 
Therefore, in this section, we show how to translate frequency-dependent accuracy requirements on the assembly model to frequency-dependent accuracy requirements on the input-to-output behavior at \emph{component}-level (i.e., how to perform assembly-aware component reduction).

\subsection{Choice of assembly model accuracy requirements}
For models of mechanical systems, we generally require the model to be accurate up to a certain frequency, often defined by a maximum frequency of interest $\omega_{\max}$. 
Additionally, models are often required to be especially accurate around specific frequency ranges, e.g., around frequencies that are present in the motion profile of a system, or around resonances that are crucial for the structural integrity of a system.
Therefore, as defined in \cite{janssen2023modular}, we introduce the assembly accuracy requirement such that the error dynamics $E_c(i\omega)$ is contained in the set
\begin{align}
\label{eq:Ec_bound}
\E_c(\omega) := \big\{ E_c(i\omega) \ \big| \ \|V_c(\omega)E_c(i\omega)W_c(\omega)\| < 1 \big\}.
\end{align}
Here, $E_c(i\omega) \in \E_c(\omega)$ is a frequency-dependent accuracy requirement that allows us to define exactly the required accuracy of the assembly model at any frequency $\omega$.
Diagonal, frequency-dependent scaling matrices $V_c(\omega) \in \R^{p_c \times {p_c}}_{>0}$ and $W_c(\omega) \in \R^{m_c \times {m_c}}_{>0}$ can be used to scale the input-output pairs of $E_c(i\omega)$ to fit the assembly accuracy requirements at the respective frequency.  As such, $W_c$ and $V_c$ reflect the assembly accuracy requirements.
%Note that $\|\cdot\|$ denotes the 2-norm of a matrix. 

%For example, in an assembly model for which we require high accuracy at 120 rad/s for the dynamics between a specific external input to a specific external output, we can choose large values for the respective diagonal elements of $V_c(120)$ and $W_c(120)$. 
%In doing so, the required error $E_c(120)$ needs to be small for that input-output pair in order to satisfy the assembly requirement $E_c(120i) \in \E_c(120)$, i.e., $\|V_c(120)E_c(120i)W_c(120)\| < 1$.
%In the next subsection, we demonstrate how this requirement can be translated to local component requirements.

\subsection{Synthesis of component model accuracy requirements}
In \cite{janssen2022modular,janssen2023modular}, it has been shown how requirements on $E_c(i\omega)$ can be used to compute component-level requirements on $E^{(j)}(i\omega)$ that, when satisfied, guarantee satisfaction of the assembly model accuracy requirements.
Using this approach as a basis, we will construct sets
\begin{align}
\label{eq:Ej_bound}
\E^{(j)}(\omega) := \big\{ E^{(j)}(\omega) \ \big| \ \| \bigl((W^{(j)}(\omega))^{-1}E^{(j)}(i\omega)(V^{(j)}(\omega))^{-1} \| \leq 1 \big\}
\end{align}
such that $E^{(j)}(i\omega) \in \E^{(j)}(\omega)$ for all $j \in \bk$ implies $E_c(i\omega) \in \E_c(\omega)$. In (\ref{eq:Ej_bound}), $V^{(j)}(\omega)$ and $W^{(j)}(\omega)$ are diagonal, frequency-dependent scaling matrices.
%In this work, we show how this allows for components to be reduced individually using CMS methods, while only having to meet local requirements on $E^{(j)}(\omega)$.
%This provides a method to guarantee the assembly requirements a priori, i.e., before computing the assembly model $\hat{\Sigma}_c$.
%In \cite{janssen2022modular}, we have shown that to find the component accuracy requirements characterized by $(\E^{(1)}(\omega),\dots,\E^{(k)}(\omega))$ based on the requirement $\E_c(\omega)$ for $\omega \in \R$, we can reformulate the problem as a robust performance problem.
Furthermore, we define
\begin{align}
\label{eq:N}
N(i\omega)&:=\left[\begin{array}{cc}
\K_{bb}(I-H_b(i\omega)\K_{bb})^{-1} & (I-\K_{bb}H_b(i\omega))^{-1} \K_{bc}\\ 
\K_{cb}(I-H_b(i\omega)\K_{bb})^{-1}&O
\end{array}\right], \\
\label{eq:setV}
\mathbf{V} &:= \Big\{\textrm{diag}(V^{(1)},\dots,V^{(k)},V_c) \Big| V^{(j)}{=}\textrm{diag}(v_j \in \R^{m_j}_{>0}) \ \forall j {\in} \bk, V_c{=}\textrm{diag}(v_c \in \R^{p_c}_{>0}) \Big\}, \\
\label{eq:setW}
\mathbf{W} &:= \Big\{\textrm{diag}(W^{(1)},\dots,W^{(k)},W_c) \Big| W^{(j)}{=}\textrm{diag}(w_j \in \R^{p_j}_{>0}) \ \forall j {\in} \bk, W_c{=}\textrm{diag}(w_c \in \R^{m_c}_{>0}) \Big\}, \\
\label{eq:setD}
\mathbf{D} &:= \Big\{ (D_\ell, D_r) \ \Big| \ d_1,\dots,d_k, d_c \in \R_{>0}, \ D_\ell = \textrm{diag}\left( d_1 I_{p_1},\dots,d_{k} I_{p_k}, d_cI_{m_c} \right), \\ \nonumber
& \qquad \qquad D_r = \textrm{diag}\left( d_1 I_{m_1},\dots,d_{k} I_{m_k}, d_cI_{p_c} \right)\Big\}.
\end{align}
Then, for any $\omega\in\R$, the following optimization problem can be solved~\cite{janssen2023modular}:
\begin{align}
\label{eq:top_down}
\textrm{given} \quad & V_c(\omega), W_c(\omega) \nonumber \\
\textrm{minimize} \quad & \tr(V^{-2}(\omega)) + \tr(W^{-2}(\omega)) \nonumber \\
\textrm{subject to} \quad & \left[\begin{array}{cc}
W^{-2}(\omega)D_r^{-1} 	& 	N^H(i\omega) \nonumber \\
N(i\omega)				& 	V^{-2}(\omega)D_\ell
\end{array}\right] \succ 0, \nonumber \\
\quad & V(\omega) \in \mathbf{V}, W(\omega) \in \mathbf{W}, (D_\ell, D_r) \in \mathbf{D}. 
\end{align}
The optimization problem (\ref{eq:top_down}) aims to ``maximize" the weights $V^{(j)}$ and $W^{(j)}$ to make the sets in (\ref{eq:Ej_bound}) as large as possible. 
This will allow for larger component reductions while maintaining the desired assembly-level accuracy.
It follows from \cite[Theorem 1]{janssen2023modular}, that for any feasible solution to (\ref{eq:top_down}), our requirement on the assembly model $E_c(i\omega) \in \E_c(\omega)$ is met if $E^{(j)}(i\omega) \in \E^{(j)}(\omega)$ as in (\ref{eq:Ej_bound}) for all $j \in \bk$.
Given the error requirements $\E^{(j)}(\omega)$, each individual component can then be reduced individually using any MOR method.
%For any ROM satisfying the component-level requirements $\E^{(j)}(\omega)$, it is guaranteed that the assembly model requirement $\E_c(\omega)$ are met.

\subsection{Proposed approach}\label{sec:proposed_approach}
Using the results from the previous section, we propose the following main approach to effectively and efficiently compute accurate ROMs of assemblies of multiple components using CMS:
\begin{enumerate}
\item Establish the initial discretized second-order models $\Sigma^{(j)}$ as (\ref{eq:sigma_j}) for all components $j \in \bk$.
\item Specify the assembly accuracy requirements by choosing $\E_c(\omega)$ as in~(\ref{eq:Ec_bound}), this includes specifying frequency of interest $\Omega_{\text{int}} := \{\omega \ | \ 0 < \omega \leq \omega_{\max} \}$ and weighting filters $W_c(\omega)$ and $V_c(\omega)$ for $\omega \in \Omega_{\text{int}}$.
\item Compute $H^{(j)}(\omega)$ for all $j \in \bk$, formulate the interconnection structure $\K$, and determine $N(\omega)$ as in (\ref{eq:N}) for $\omega \in \Omega_{\text{int}}$.
\item Solve the optimization problem (\ref{eq:top_down}) to find $\E^{(j)}$ as in (\ref{eq:Ej_bound}) for all $j \in \bk$. Note that with (\ref{eq:top_down}), the requirements $\E^{(j)}$ are distributed such that components that are relatively more important to the overall behaviour of the assembly model are assigned stricter requirements in comparison to less important components \cite{janssen2023modular}.
\item Given these requirements for all components $j \in \bk$, independently compute reduced-order component models $\hat{\Sigma}^{(j)}$ using any suitable CMS method (or reduction method in general) by taking the minimum modes $\hat{n}_j$ required to meet its component requirement $\E^{(j)}$. Note that this is computationally relatively cheap, since the satisfaction of the accuracy requirements is only checked on component-level.
\item The reduced-order assembly model $\hat{\Sigma}_c$ (built with $\hat{\Sigma}^{(j)}$ for all $j \in \bk$ and $\K$) is now guaranteed to meet the given requirement $\E_c(\omega)$.
\end{enumerate}
Note that in practice, to reduce the computational cost of the approach, in step (1), we can regard as $\Sigma^{(j)}$ the models obtained by applying HH reduction to the original discretization. Here, we can include all modes up to many times the maximum frequency of interest, e.g., 10 times $\omega_{\max}$.
Additionally, frequencies $\omega$ are typically selected in a grid on $\Omega_{\text{int}}$.
%In the next section, the approach is applied to an example system.

\section{ILLUSTRATIVE EXAMPLE}
\label{sec:example}
\begin{figure}
\begin{subfigure}{.5\textwidth}
  	\centering
   	\includegraphics[scale=0.18]{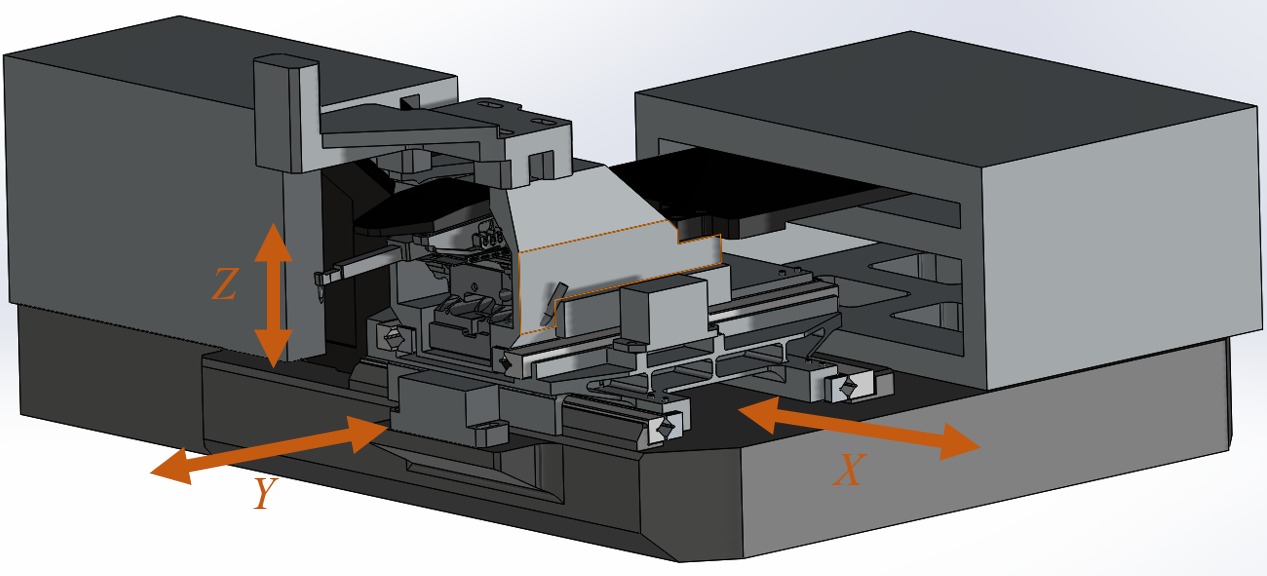}
%   	\vspace{1pt}
   	\caption{3D Industrial model}
    \label{fig:wirebonder_pic}
\end{subfigure}
\label{fig:wirebonder}
\begin{subfigure}{.5\textwidth}
  	\centering
   	\includegraphics[trim={.8cm 1cm 1.6cm 2.2cm},clip, scale=0.55]{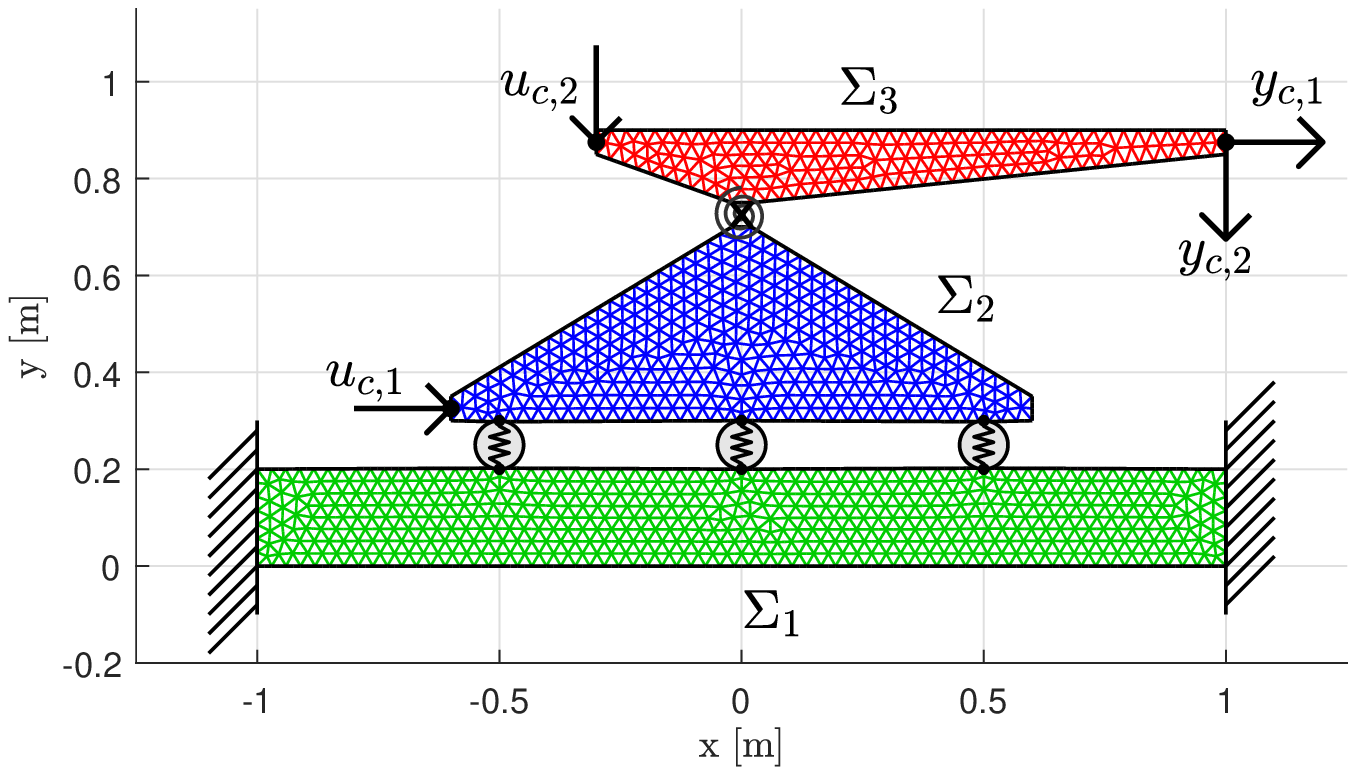}
   	\vspace{-8pt}
   	\caption{2D Finite Element model}
    \label{fig:wirebonder_example}
\end{subfigure}
\vspace{-10pt}
\caption{Industrial and simplified model of a wirebonder machine.}
\end{figure}

In this section, we apply the proposed approach on a 2D structural dynamics FE assembly model, inspired by a 3D industrial wire bonder system (Figure~\ref{fig:wirebonder_pic}). The FE assembly model consists of three elastically interconnected components (Figure~\ref{fig:wirebonder_example}).
In principle, the system in Figure~\ref{fig:wirebonder_example} allows for 1) motion in the horizontal direction using bearings (modeled by three translational springs) between the machine frame ($\Sigma_1$) and the y-stage ($\Sigma_2$) and 2) rotational motion of the z-stage ($\Sigma_3$) about the y-stage via a cross leaf-spring. 
However, to demonstrate our methodology, we will linearize the model around one working point.
To compute ROMs of this linear structural dynamics system, we follow the approach of Section~\ref{sec:proposed_approach}.

\subsection{System description}
All three-component FE models have a Young's modulus of $210\times 10^{9}$ Pa, a Poisson ratio of $0.3$ and a mass density of 7800 kg/m$^3$ and are discretized with quadratic triangular plane-stress elements. All components have $1\%$ modal damping.
\begin{enumerate}
\item The machine frame model, $\Sigma^{(1)}$, is a rectangular body that is attached to the fixed-world at its left and right end, and it is connected at its top to the y-stage through three springs (each with stiffness $k_b = 2$ GN/m) that represent the vertical stiffness in the bearings.
\item The y-stage, $\Sigma^{(2)}$, is a triangular body that has a horizontal external force $u_{c,1}$ applied to its left end. Furthermore, it is attached at its top to the z-stage through a cross leaf-spring connection with a high stiffness $k_{t} = 20$ GN/m in the horizontal and vertical directions, and a low stiffness $k_{r} = 0.5$ GNm/rad in the rotational direction.
\item The z-stage, $\Sigma^{(3)}$, is a triangular body with a vertical external force $u_{c,2}$ applied to its left end. Furthermore, the point-of-interest (POI) of the system is the right end of the z-stage, for which the horizontal and vertical positions are measured by outputs $y_{c,1}$ and $y_{c,2}$, respectively.
\end{enumerate}
For the unreduced assembly model, Equation (\ref{eq:sigmac}) is then given by
\begin{align}
\left[\begin{array}{c}
u^{(1)} \\
u^{(2)} \\
u^{(3)} \\
y_c
\end{array}\right] = \left[\begin{array}{ccc|cccccc|ccccc|cc}
\vspace{-2pt}
\hspace{-5pt}-k_b\hspace{-5pt} & & & k_b & & & & & & & & & & & & \\
\vspace{-2pt}
& \hspace{-5pt}-k_b\hspace{-5pt} & & & k_b & & & & & & & & & & & \\
\vspace{-2pt}
& & \hspace{-5pt}-k_b\hspace{-5pt} & & & k_b & & & & & & & & & & \\
\hline
\vspace{-2pt}
k_b & & & \hspace{-5pt}-k_b\hspace{-5pt} & & & & & & & & & & & & \\
\vspace{-2pt}
& k_b & & & \hspace{-5pt}-k_b\hspace{-5pt} & & & & & & & & & & & \\
\vspace{-2pt}
& & k_b & & & \hspace{-5pt}-k_b\hspace{-5pt} & & & & & & & & & & \\
\vspace{-2pt}
& & & & & & \hspace{-5pt}-k_t\hspace{-5pt} & & & k_t & & & & & & \\
\vspace{-2pt}
& & & & & & & \hspace{-5pt}-k_t\hspace{-5pt} & & & k_t & & & & & \\
\vspace{-2pt}
& & & & & & & & \hspace{-5pt}-k_r\hspace{-5pt} & & & k_r & & & & \\
\vspace{-2pt}
& & & & & & & & & & & & & & 1 & \\
\hline
& & & & & & k_t & & & \hspace{-5pt}-k_t\hspace{-5pt} & & & & & & \\
\vspace{-2pt}
& & & & & & & k_t & & & \hspace{-5pt}-k_t\hspace{-5pt} & & & & & \\
\vspace{-2pt}
& & & & & & & & k_r & & & \hspace{-5pt}-k_r\hspace{-5pt} & & & & \\
\vspace{-2pt}
& & & & & & & & & & & & & & & 1 \\
\hline
& & & & & & & & & & & & 1 & & & \\
\vspace{-2pt}
& & & & & & & & & & & & & 1 & & \\
\end{array}\right]\hspace{-5pt}
\left[\begin{array}{c}
y^{(1)} \\
y^{(2)} \\
y^{(3)} \\
u_c
\end{array}\right]\hspace{-5pt}.
\end{align}

\subsection{Accuracy requirements $\E_c$ and $\E^{(j)}$}
\begin{figure}
  	\centering
   	\includegraphics[trim={1cm 0cm 1cm .3cm},clip, scale=1]{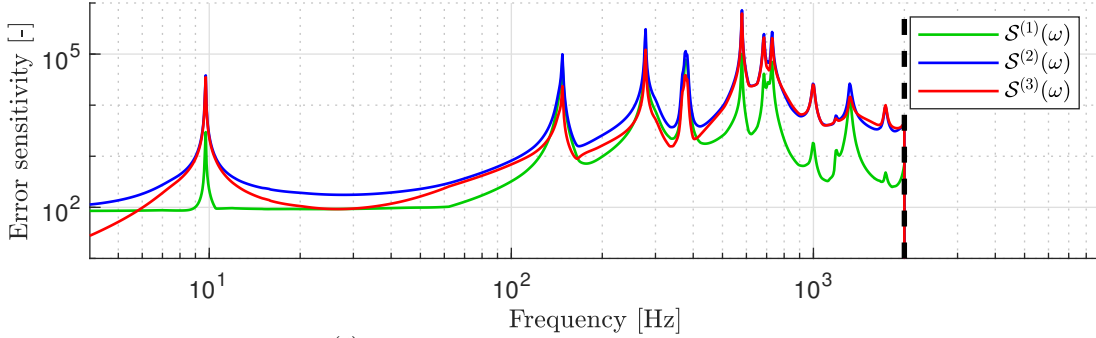}
   	\vspace{-10pt}
   	\caption{Error sensitivity $\mathcal{S}^{(j)}(\omega)$ for each subsystem $j \in \bk$ and for $\omega \in \Omega_{\text{int}}$ that measures how quickly $E^{(j)}(i\omega)$ exceeds $\E^{(j)}(\omega)$, as a solution to (\ref{eq:top_down}), given $\E_c$.}
    \label{fig:result_Rel}
\end{figure}

For this system, the maximum frequency of interest is $f_{\max} := \omega_{\max}/(2\pi) = 2000$ Hz and the assembly model accuracy requirements are based on a maximum on the relative error (of 5\%), i.e., 
\begin{align}
\label{eq:ex_req}
\frac{\|E_c(i\omega)\|}{\|H_c(i\omega)\|} = \frac{\|\hat{H}_c(i\omega) - H_c(i\omega)\|}{\|H_c(i\omega)\|} < \gamma = 0.05
\end{align}
for all $\omega \in \Omega_{\text{int}}$. 
We have that $E_c(i\omega) \in \E_c(\omega)$ as in (\ref{eq:Ec_bound}) guarantees (\ref{eq:ex_req}) for the choice $V_c(\omega) = W_c(\omega) = I_{2}(\gamma\|H_c(i\omega)\|)^{-\frac{1}{2}}$.
Given these assembly model accuracy requirement $\E_c(\omega)$ for $\omega \in \Omega_{\text{int}}$, the optimization problem (\ref{eq:top_down}) can be solved following the approach described in Section~\ref{sec:proposed_approach}.
The unique solution will provide component accuracy requirements $\E^{(j)}(\omega)$ for all $j \in \bk$.
To compare the component accuracy requirements, we introduce a measure of sensitivity $\mathcal{S}(\omega)$, that measures, for $\omega \in \Omega_{\text{int}}$, how quickly component errors $E^{(j)}(i\omega)$ exceed $\E^{(j)}(\omega)$, given by $\mathcal{S}^{(j)}(\omega) :=\|(W^{(j)}(\omega))^{-1}J(V^{(j)}(\omega))^{-1}\|$.
Since $V^{(j)}$ and $W^{(j)}$ are not necessarily of the same dimension, $J$ is a matrix of ones of the appropriate dimensions.
The resulting error sensitivity for each component is given in Figure~\ref{fig:result_Rel}.
From this figure, it becomes clear that the assembly model accuracy is less sensitive to errors in component 1, i.e., the machine frame, which means that a higher reduction error is allowed.

\subsection{Model reduction: standard vs. proposed approach}
\begin{figure}
  	\centering
   	\includegraphics[trim={1cm 0cm 1cm .3cm},clip, scale=1]{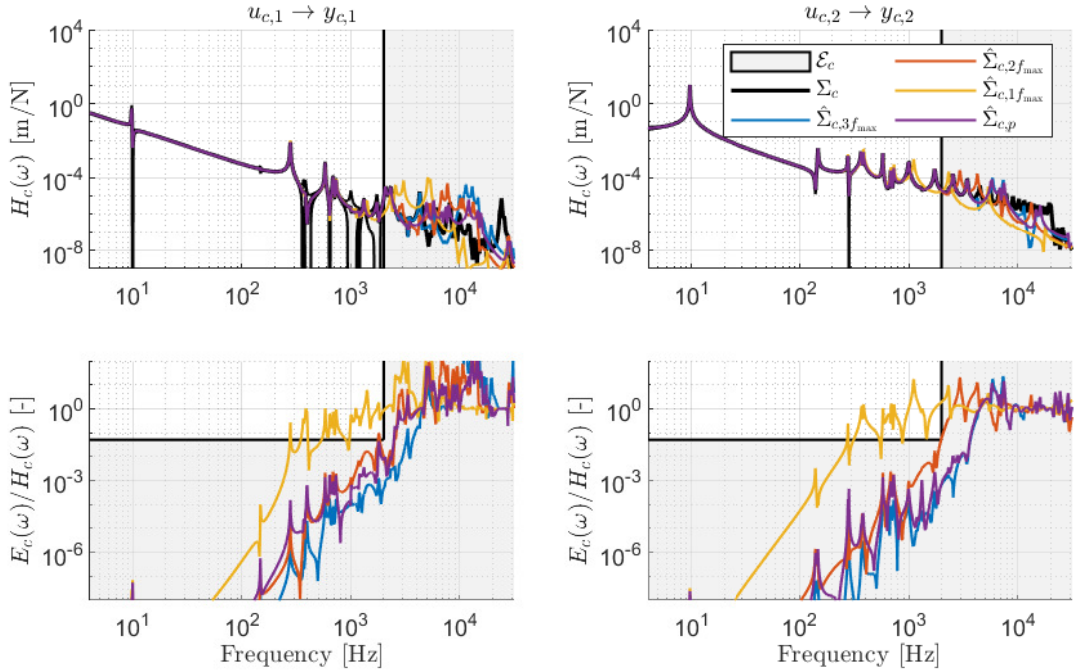}
   	\vspace{-7pt}
   	\caption{Two FRFs of the different ROMs (on top) and the corresponding relative errors (on the bottom) for the different ROMs, including the error requirement $\E_c$.}
    \label{fig:Gc}
\end{figure}

In this example, we regard as the high-order assembly model $\Sigma_c$ the model obtained by applying HH to the original discretization by including all modes up to $10\omega_{\max}$ (262 modes in total).
Furthermore, two methods will be compared:
\begin{enumerate}
\item The \emph{standard} ROMs $\hat{\Sigma}_{c,i f_{\max}}$ for $i = 1,2,3$ are computed by applying HH and including all modes up to three predefined cut-off frequencies $f_\cut = i f_{\max}$ with $f_{\max} = 2000$ Hz, which are the same for each component. 
\item The \emph{proposed} ROM $\hat{\Sigma}_{c,p}$ is computed using the approach proposed in Section~\ref{sec:proposed_approach}.%, i.e., by computing the ROMs $\hat{\Sigma}^{(j)}_p$ with the lowest number of DOF satisfying $E^{(j)}(\omega) \in \E^{(j)}(\omega)$ on component-level for all $j \in \bk$. 
\end{enumerate}
The number of modes required to meet the requirement for these ROMs are given in Table~\ref{tab:results} and the corresponding FRFs and relative error for the different ROMs are given in Figure~\ref{fig:Gc}.
%\begin{table}[]
%\caption{Reduction results for the four ROMs with the component cut-off frequency $f_\cut^{(j)}$, requirement satisfaction $\E^{(j)} \in E^{(j)}$ and number of modes $\hat{n}_j$ for each component $j = 1,2,3$, and the requirement satisfaction $\E_c \in E_c$ and number of modes $\hat{n}_c$ for the assembly model.}
%\label{tab:results}
%\centering
%\begin{tabular}{ll|lllll}
%Assembly model & & $\Sigma_c$ & $\hat{\Sigma}_{c,3f_{\max}}$ & $\hat{\Sigma}_{c,2f_{\max}}$ & $\hat{\Sigma}_{c,1f_{\max}}$ & $\hat{\Sigma}_{c,p}$ \\
%\hline
%Cut-off frequency [Hz] & $f_{\cut}^{(1)}$  & 20,000 & 6,000 & 4,000 & 2,000 & 1,976 \\
%& $f_{\cut}^{(2)}$ & 20,000 & 6,000 & 4,000 & 2,000 & 3,956\\
%& $f_{\cut}^{(3)}$ & 20,000 & 6,000 & 4,000 & 2,000 & 5,768\\
%\hline
%\# of modes & $\hat{n}_1$ & 106 & 16 & 12 & 8 & 6 \\
%& $\hat{n}_2$ & 94 & 18 & 14 & 11 & 13\\
%& $\hat{n}_3$ & 62 & 17 & 14 & 10 & 16\\
%& $\hat{n}_c$ & \textbf{262} & \textbf{51} & \textbf{40} & \textbf{29} & \textbf{35}\\
%\hline
%Requirement satisfaction & $E^{(1)}\in\E^{(1)}$ & \green{yes} & \green{yes} & \green{yes} & \green{yes} & \green{yes}\\
%& $E^{(2)}\in\E^{(2)}$ & \green{yes} & \green{yes} & \green{yes} & \red{no} & \green{yes}\\
%& $E^{(3)}\in\E^{(3)}$ & \green{yes} & \green{yes} & \red{no} & \red{no} & \green{yes}\\
%& $E_c\in\E_c$ & \green{\textbf{yes}} & \green{\textbf{yes}} & \red{\textbf{no}} & \red{\textbf{no}} & \green{\textbf{yes}}\\
%\end{tabular}
%\end{table}

From these results, several conclusions and observations can be made.
First, in Table~\ref{tab:results}, we observe that $\hat{\Sigma}_{c,3f_{\max}}$ satisfies the accuracy requirements $E^{(j)} {\in} \E^{(j)}$ for all components. Therefore, $\hat{\Sigma}_{c,3f_{\max}}$ satisfies the assembly requirement $E_c {\in} \E_c$.
In contrast, $\hat{\Sigma}_{c,2f_{\max}}$ satisfies $E^{(j)} {\in} \E^{(j)}$ only for components 1 and 2, $\hat{\Sigma}_{c,1f_{\max}}$ only for component 1. 
As a result, $\hat{\Sigma}_{c,2f_{\max}}$ and $\hat{\Sigma}_{c,1f_{\max}}$ do not satisfy the assembly requirement $E_c {\in} \E_c$.
In comparison, the proposed $\hat{\Sigma}_{c,p}$ satisfies (by definition), the requirements for all components, therefore $E_c {\in} \E_c$ is guaranteed.
Because for $\hat{\Sigma}_{c,p}$ the cut-off frequency is chosen adaptively, i.e., by minimizing the number of modes satisfying $E^{(j)} {\in} \E^{(j)}$, components 1 and 2 can be reduced further with respect to $\hat{\Sigma}_{c,3f_{\max}}$ and $\hat{\Sigma}_{c,2f_{\max}}$ and component 3 needs to be relatively accurate to meet the requirements.
In Figure~\ref{fig:Gc}, we can observe that the FRFs for the $\hat{\Sigma}_{c,1f_{\max}}$ and $\hat{\Sigma}_{c,2f_{\max}}$ indeed exceed the assembly requirements and $\hat{\Sigma}_{c,3f_{\max}}$ and $\hat{\Sigma}_{c,p}$ satisfy them.
\begin{table}[]
\caption{Comparison between models resulting from the standard approach $\hat{\Sigma}_{c,1f_{\max}}$, $\hat{\Sigma}_{c,2f_{\max}}$ and $\hat{\Sigma}_{c,3f_{\max}}$, and from the proposed approach $\hat{\Sigma}_{c,p}$. Reduction results for the ROMs with the component cut-off frequency $f_\cut^{(j)}$ in Hz, number of modes $\hat{n}_j$ and requirement satisfaction $\E^{(j)} \in E^{(j)}$ for each component $j = 1,2,3$, and the number of modes $\hat{n}_c$ and requirement satisfaction $\E_c \in E_c$ for the assembly model.}
\label{tab:results}
\vspace{-8pt}
\begin{tabular}{l|lll|lll|lll|ll}
 & \multicolumn{3}{l|}{Subsystem 1} & \multicolumn{3}{l|}{Subsystem 2} & \multicolumn{3}{l|}{Subsystem 3} & \multicolumn{2}{l}{Assembly} \\
 & $f_{\cut}^{(1)}$ & $\hat{n}_1$ & $E^{(1)}{\in}\E^{(1)}$\hspace{-5pt} & $f_{\cut}^{(2)}$ & $\hat{n}_2$ & $E^{(2)}{\in}\E^{(2)}$\hspace{-5pt} & $f_{\cut}^{(3)}$ & $\hat{n}_3$ & $E^{(3)}{\in}\E^{(3)}$\hspace{-5pt} & $\hat{n}_c$ & $E_c{\in}\E_c$\hspace{-5pt} \\
 \hline
$\Sigma_c$ & 20,000\hspace{-5pt} & 106 & \green{yes} & 20,000\hspace{-5pt} & 94 & \green{yes} & 20,000\hspace{-5pt} & 62 & \green{yes} & \textbf{262} & \green{\textbf{yes}} \\
 \hline
$\hat{\Sigma}_{c,3f_{\max}}$\hspace{-5pt} & 6,000 & 16 & \green{yes} & 6,000 & 18 & \green{yes} & 6,000 & 17 & \green{yes} & \textbf{51} & \green{\textbf{yes}} \\
$\hat{\Sigma}_{c,2f_{\max}}$\hspace{-5pt} & 4,000 & 12 & \green{yes} & 4,000 & 14 & \green{yes} & 4,000 & 14 & \red{no} & \textbf{40} & \red{\textbf{no}} \\
$\hat{\Sigma}_{c,1f_{\max}}$\hspace{-5pt} & 2,000 & 8 & \green{yes} & 2,000 & 11 & \red{no} & 2,000 & 10 & \red{no} & \textbf{29} & \red{\textbf{no}} \\
 \hline
$\hat{\Sigma}_{c,p}$ & 1,976 & 6 & \green{yes} & 3,956 & 13 & \green{yes} & 5,768 & 16 & \green{yes} & \textbf{35} & \green{\textbf{yes}}
\end{tabular}
\end{table}
Summarizing, the reduced-order assembly model $\hat{\Sigma}_{c,p}$ resulting from our new approach benefits from the fact that component-level accuracy requirements become available.
Selecting the lowest cut-off frequency per component that meets these requirements, results in a reduced assembly model that   
\begin{enumerate}
\item is of much lower order (35 modes) compared to the only ROM computed with the standard approach that satisfies the assembly requirements, i.e., $\Sigma_{c,f_{\max}}$ (51 modes), which means that the computational cost to apply the assembly model in simulation is lower, and
%\item is of lower order in comparison more between taking a fixed cut-off frequency 2 times the maximum frequency of interest (40 total modes), even though this model does not even satisfy the assembly requirements, and 
\item is guaranteed to satisfy accuracy requirements on the assembly model, even though the ROMs are computed and verified on component-level only.
\end{enumerate}
 
\section{CONCLUSIONS}
\label{sec:conclusions}
In this work, we demonstrate how accuracy requirements on an assembly model can be translated to accuracy requirements on the level of its components.
This translation allows for the independent reduction of components for which only the component-level accuracy requirement needs to be satisfied in order to guarantee the final accuracy of the reduced-order assembly model.
Specifically, it provides a method to find an individual cut-off frequency for each component.
This has the advantage that 1) components can be reduced based on their relative importance to the behavior of the complete assembly model and 2) components may be reduced further in comparison to the standard approach of taking all modes up to a fixed, a priori chosen, cut-off frequency.
The approach is applied, using the Hintz-Herting reduction method, to a simplified mechanical model of a wirebonder system.
In this example, it was shown that the system model could be reduced further in comparison to the standard approach, while guaranteeing that the assembly model satisfies the given accuracy requirements.

\section{ACKNOWLEDGEMENTS}
This publication is part of the project Digital Twin with project number P18-03 of the research programme Perspectief which is (mainly) financed by the Dutch Research Council (NWO).

\bibliography{eurodyn}  

\end{document}